%% file: draft_single.tex
\documentclass[12pt,journal,draftcls,a4paper,onecolumn]{IEEEtran} 
\usepackage{dsfont}

\usepackage{subeqnarray}
\usepackage{cleveref}
\usepackage{amssymb}
\usepackage{amsfonts}
\usepackage{amsmath}
\usepackage[dvipdfm]{graphicx}
\usepackage{bmpsize}
\usepackage{amsmath}
\usepackage[all,poly]{xy}
\usepackage{multirow}
\usepackage{algorithm}
\usepackage{algorithmic}
\usepackage{color}
\usepackage[noadjust]{cite}

\title{Lidar waveform based analysis of depth images constructed using sparse single-photon data}
\author{Yoann Altmann, Ximing Ren, Aongus McCarthy, Gerald S. Buller and Steve
McLaughlin 
\thanks{Yoann Altmann, Ximing Ren, Aongus McCarthy, Gerald Buller and Steve
McLaughlin are with School of Engineering and Physical Sciences, Heriot-Watt University,
U.K. (email: \{Y.Altmann;xr5;A.McCarthy; G.S.Buller;S.McLaughlin\}@hw.ac.uk).}
\thanks{This study was supported by EPSRC via grants EP/J015180/1, EP/K015338/1 and EP/M01326X/1.
}}

\newcommand{\bGam}{\boldsymbol{\Gamma}}

\input notations_yoann.tex

\begin{document}
\maketitle

\begin{abstract}
This paper presents a new Bayesian model and algorithm used for depth 
and intensity profiling using full waveforms from the time-correlated 
single photon counting (TCSPC) measurement in the limit of very low photon counts. The
model proposed represents each Lidar waveform as a combination 
of a known impulse response, weighted by the target intensity, 
and an unknown constant background, corrupted by Poisson noise. Prior
knowledge about the problem is embedded in a hierarchical
model that describes the dependence structure between the model
parameters and their constraints. In particular, a gamma Markov
random field (MRF) is used to model the joint distribution of the
target intensity, and a second MRF is used to model the distribution of the target depth, which are both expected to exhibit significant spatial
correlations. An adaptive Markov chain Monte Carlo algorithm is
then proposed to compute the Bayesian estimates of interest and
perform Bayesian inference. This algorithm is equipped with a
stochastic optimization adaptation mechanism that automatically
adjusts the parameters of the MRFs by
maximum marginal likelihood estimation. Finally, the benefits of the proposed
methodology are demonstrated through a serie of experiments
using real data.
\end{abstract}


\section{Introduction}
Reconstruction of 3-dimensional scenes using time-of-flight laser imaging detection and ranging (Lidar) systems is a challenging problem encountered in many applications, including automotive \cite{Ogawa2006,Lindner2009,Matzka2012,Yeonsik2012}, environment sciences \cite{Ramirez2012,Hakala2012}, architectural engineering and defence \cite{Cadalli2002,Gao2011}. This problem consists of illuminating the scene with a train of laser pulses and analyzing the distribution of the photons reflected from the targets to infer the range, as well as intensity information about the scene. Using scanning systems, an histogram of time delays between the emitted pulses and the detected photon arrivals is usually recorded for each pixel, associated with a different region of the scene. The recorded photon histograms are classically decomposed into a series of peaks whose positions can be used to infer the distance of the object(s) present in each region of the scene and whose amplitudes provide information about the intensity of the objects. 

In this paper, we consider applications where the flux of detected photons is small and for which classical methods usually provide unsatisfactory results in terms of range and intensity estimation. This is typically the case when the acquisition time or the laser source power are small relative to the range of the target(s) \cite{McCarthy2013}. 

Recently, Kirmani \emph{et al.} \cite{Kirmani2014}, investigated a new method for reconstructing 3-dimensional scenes under low photon flux conditions by registering the time of arrival of the first photon in each pixel. Based on an appropriate statistical model relating the time of arrival to the target distance and intensity, they proposed different processing steps, to handle ambient noise (background photons) and the spatial regularity of the scene to obtain estimated target distances and intensities. In contrast with the method proposed in \cite{Kirmani2014}, we consider a scanning system whose acquisition time per pixel is fixed, thus avoiding any potential synchronisation issues and leading to
a deterministic and user-defined overall acquisition duration. Consequently, the number of detected photons can be larger than one for some pixels, whereas some pixels may be empty, i.e., contain no detected photons. 

In this work, we assume that the targets of interest are opaque, i.e., are composed of a single surface per pixel and that at least a surface is present in each scanned pixel. Generalization to more complex objects will be discussed in the conclusion of this paper. As in \cite{Kirmani2014}, we consider the presence of two kinds of detector events: the "useful" photons originating from the illumination laser and scattered back from the target; and those background detector events originating from ambient light and the "dark" events resulting from detector noise. The proposed method aims to estimate the respective contributions of the actual target and the background in the photon timing histograms. More precisely, it also allows the estimation of the distance and intensity of the surface associated with each pixel, together with the average background levels, within a single estimation procedure.  

Adopting a Bayesian framework as in \cite{Hernandez2007,Wallace2014}, we assign prior distributions to the unknown model
parameters to include available information (such as parameter constraints) within the estimation procedure. In particular, Markov random fields (MRFs) are introduced to model
spatial correlations for the target distances and intensities. The joint posterior distribution of the
unknown parameter vector is then derived, based on the Poisson statistical properties of the observed data. Since classical Bayesian estimators cannot be
easily computed from this joint posterior, a Markov chain Monte Carlo (MCMC) method is
used to generate samples according to this posterior. More precisely, we construct an efficient
stochastic gradient MCMC (SGMCMC) algorithm \cite{Pereyra2014ssp} that simultaneously estimates the background levels and the target distances and intensity, along with the MRFs parameters.

The main contributions of this work are threefold:
\begin{enumerate}
\item  We develop new hierarchical intensity and depth models taking into account spatial
correlations through Markovian dependencies. These flexible models are embedded within the observation model for full waveform Lidar-based low photon count imaging

\item  An adaptive Markov chain Monte Carlo algorithm is then proposed to compute the
Bayesian estimates of interest and perform Bayesian inference. This algorithm is equipped
with a stochastic optimization adaptation mechanism that automatically adjusts the
parameters of the Markov random fields by maximum marginal likelihood estimation,
thus removing the need to set the regularization parameters by cross-validation.
\item We show the benefits of the proposed flexible model for reconstructing a real 3D object in scenarios where the number of detected photons is very low. 
Specifically, we demonstrate the ability of
the proposed algorithm to handle empty pixels as well as background levels.
\end{enumerate}

The remainder of this paper in organized as follows. Section \ref{sec:model} recalls the classical statistical model used for depth imaging using time-of-flight scanning sensors, based on TCSPC. Section \ref{sec:bayesian} presents the proposed hierarchical Bayesian model for low photon count depth imaging, which takes into accounts in the inherent spatial correlations between pixels. The estimation of the Bayesian model parameters using adaptive MCMC methods is discussed in Section \ref{sec:Gibbs}. Simulation results conducted using an actual time-of-flight scanning sensor are presented and discussed in Section \ref{sec:simulations}. Finally, conclusions and potential future work are reported in Section \ref{sec:conclusion}.

\section{Problem formulation}
\label{sec:model} We consider a set of $N_{\textrm{row}} \times N_{\textrm{col}}$ observed Lidar waveforms/pixels 
$\Vpix{i,j} = [\pix{i,j}{1},\ldots,\pix{i,j}{\nbbin}]\transp, (i,j) \in \{1,\ldots,N_{\textrm{row}}\} \times \{1,\ldots,N_{\textrm{col}}\}$ where $\nbbin$ is the number of temporal
(corresponding to range) bins.
To be precise, $\pix{i,j}{\nobin}$ is the photon count within the $\nobin$th bin of the pixel or location $(i,j)$. Let $\nobin_{i,j}$ be the position of an object surface at a given range from the sensor. According to \cite{Hernandez2007}, each photon count $\pix{i,j}{\nobin}$ is assumed to be drawn from the following Poisson distribution 
\begin{eqnarray}
\label{eq:model0}
\pix{i,j}{\nobin} \sim \mathcal{P}\left(r_{i,j} g_{0}\left(\nobin-\nobin_{i,j}\right) + b_{i,j}\right)
\end{eqnarray}
where $g_{0}(\cdot)>0$ is the photon impulse response , $r_{i,j}$ denotes the intensity of the target and $b_{i,j}$ stands for the background and dark photon level, which is constant in all bins of a given pixel.
The photon impulse response $g_{0}(\cdot)$ is assumed to be known and estimated during the imaging system calibration. 

Due to physical considerations, the target intensity is assumed to satisfy the following positivity constraints 
\begin{eqnarray}
\label{eq:pos_const}
r_{i,j} \geq 0, \quad \forall i,j.
\end{eqnarray} 
Similarly, the average background level in pixel satisfies $b_{i,j}\geq 0, \forall i,j$. The problem addressed in this paper consists of estimating the position (\emph{i.e.}, $\nobin_{i,j}$) of the targets, their intensity $r_{i,j}$ and the background levels $b_{i,j}$ from the observed data gathered in the $N_{\textrm{row}} \times N_{\textrm{col}} \times T$ array $\MATpix$. The next section studies the proposed Bayesian model to estimate the unknown parameters in \eqref{eq:model0} while ensuring the positivity constraints mentioned above.

\section{Bayesian model}
\label{sec:bayesian}
The unknown parameter vector associated with \eqref{eq:model0} contains the surfaces intensity $\bfR$, the position of the target $\bfT$ and the background levels $\bfB$ (satisfying positivity constraints), where $\left[\bfR\right]_{i,j}=r_{i,j}$, $\left[\bfT\right]_{i,j}=t_{i,j}$ and $\left[\bfB\right]_{i,j}=b_{i,j}$. This section summarizes the likelihood and the parameter priors associated with these parameters. 

\subsection{Likelihood}
\label{subsec:likelihood}
Assuming the Poisson noise realizations in all bins and for all wavelengths are independent, Eq. \eqref{eq:model0} leads to 
\begin{eqnarray}
\label{eq:likelihood}
P(\MATpix|\bfR,\bfB,\bfT) = \prod_{i,j} \prod_{\nobin=1}^{\nbbin} \dfrac{\lambda_{i,j,\nobin}^{\pix{i,j}{\nobin}}}{\pix{i,j}{\nobin}!} \exp^{-\lambda_{i,j,\nobin}}
\end{eqnarray}
where $\lambda_{i,j,\nobin}=r_{i,j}g_{0}(\nobin-\nobin_{i,j}) + b_{i,j}$. 

\subsection{Prior for the target positions}
\begin{figure}[h!]
  \centering
  \includegraphics[width=\columnwidth]{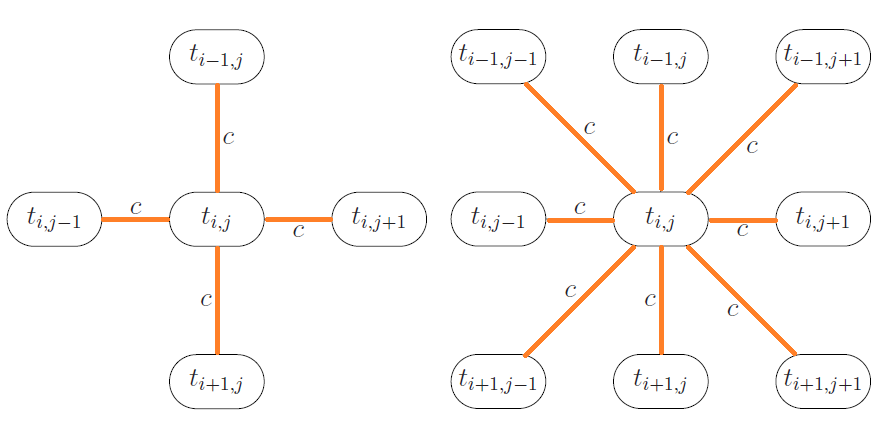}
  \caption{4-pixel (left) and 8-pixel (right) neighborhood structures. The pixel 
considered appears as a black circle whereas its neighbors are
depicted in white.}
  \label{fig:neighborhood}
\end{figure}

To account for the spatial correlations between the target distances in neighboring pixels, we propose to use a Markov random field as a prior distribution for $\nobin_{i,j}$ given its neighbors
$\bfT_{\mathcal{V}({i,j})}$, i.e., $f(t_{i,j}|\bfT_{\backslash i,j})=f(t_{i,j}|\bfT_{\mathcal{V}(i,j)})$, where $\mathcal{V}(i,j)$
is the neighborhood of the pixel $(i,j)$ and $\bfT_{\backslash i,j}=\{t_{i',j'}\}_{ (i',j')\neq (i,j)}$. More precisely, this
paper focuses on the following discrete MRF 

\begin{eqnarray}
\label{eq:MRF_depth}
f(\bfT|c) =\dfrac{1}{G(c)} \exp \left[- c \phi (\bfT) \right], \quad \forall t_{i,j} \in \{1,\ldots,T\},
\end{eqnarray}
where $c\leq 0$ is a parameter tuning the amount of correlation between pixels, $G(c)$ is a normalization (or partition) constant and where 
$\phi(\cdot)$ is an arbitrary cost function modeling the correlation between neighbors. In this work we propose to use the following cost function 
\begin{eqnarray}
\phi (\bfT) = \sum_{i,j} \sum_{(i',j')\in \mathcal{V}(i,j)}|t_{i,j}-t_{i',j'}|,
\end{eqnarray}
which corresponds to a \emph{total-variation} (TV) regularization \cite{Rudin1992,Chambolle2004} promoting piecewise constant depth image. Moreover, the higher the value of $c$, the more correlated the neighboring pixels. Several neighborhood structures can be employed to define $\mathcal{V}(i,j)$. Fig. \ref{fig:neighborhood} shows two examples of neighborhood structures. Here, the eight pixel structure (or 2-order neighborhood) will be considered in the rest of the paper for the depth parameters, as it provides in practice smoother depth images. Different cost functions (e.g., quadratic functions) could also have been used to model the depth correlations, depending on the application. Since the depths are assumed to take a finite number of values, it would not significantly change the estimation procedure presented in Section \ref{sec:Gibbs}.

\subsection{Prior for the target intensity}
In the absence of background, i.e., if $b_{i,j}=0$, gamma distributions are conjugate priors for each intensity parameter $r_{i,j}$. Consequently, it seems natural to consider gamma priors for the intensity. As will be shown later, gamma priors are still conjugate distributions when $b_{i,j}>0$.
We propose to assign $r_{i,j}$ the following gamma prior
\begin{eqnarray}
\label{eq:prior_r}
r_{i,j} \sim \mathcal{G}\left(\alpha_0,\dfrac{\alpha_{i,j}}{\alpha_0} \right)
\end{eqnarray} 
where $\alpha_{i,j}>0$ is a local parameter related to the prior mean and mode of $r_{i,j}$ and $\alpha_0>0$ is a global parameter which controls the shape of the distribution tails and thus the prior deviation of $r_{i,j}$ from $\alpha_{i,j}$. A key feature of hierarchical models is their capacity to encode
dependence and act as pooling mechanisms that share information across covariates to improve the inference. Here we specify \eqref{eq:prior_r} to reflect the prior belief that intensity exhibit
spatial correlations. In particular, due to the spatial organization of images, we expect the
values of $r_{i,j}$ to vary smoothly from one pixel to another. In order to model this behaviour, 
we specify $\alpha_{i,j}$ such that the resulting prior for $\bfR$ is a hidden gamma-MRF (GMRF) \cite{Dikmen2010}.

\begin{figure}[ht]
\centering
\includegraphics[width=\columnwidth]{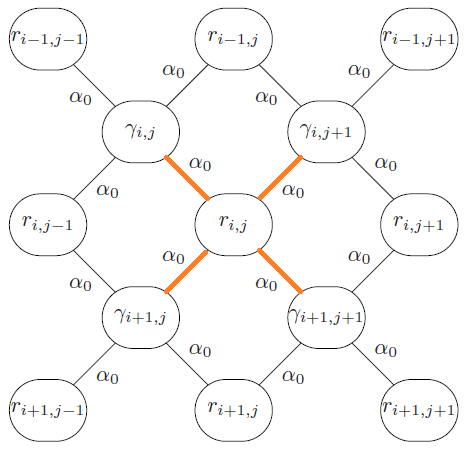}
\caption{Proposed $1$st order GMRF neighborhood structure $\forall (i,j) \in \mathcal{V}_{\bfR}$. We set $r_{i,j}=0.1, \forall (i,j) \notin \mathcal{V}_{\bfR}$}
\label{fig:neighbour_GMRF}
\end{figure}

More precisely, we introduce an $(N_{\textrm{row}}+1) \times (N_{\textrm{col}}+1)$ auxiliary matrix $\bGam$ with elements $\gamma_{i,j} \in \mathbb{R}^+$ and define a bipartite conditional independence graph between $\bfR$ and $\bGam$ such that each $r_{i,j}$ is connected to four neighbor elements of $\bGam$ and vice-versa. This $1$st order neighbourhood structure is depicted in Fig. \ref{fig:neighbour_GMRF}, where we notice that any given $r_{i,j}$ and $r_{i+1,j}$ are $2$nd order neighbors via $\gamma_{i+1,j}$ and $\gamma_{i+1,j+1}$. We specify a GMRF prior for $\bfR,\bGam$ \cite{Dikmen2010}, and obtain the following joint prior for $\bfR,\bGam$ 
\begin{eqnarray}
\label{eq:GMRF}
f(\bfR,\bGam|\alpha_0) & = & \dfrac{1}{G(\alpha_0)} \prod_{(i,j) 
\in \mathcal{V}_{\bfR}} r_{i,j}^{\left(\alpha_0-1 \right)}\nonumber\\
& \times & \prod_{(i',j') \in \mathcal{V}_{\bGam}} 
\left(\gamma_{i',j'}\right)^{-\left(\alpha_0+1 \right)} \nonumber\\
 & \times & \prod_{\left((i,j),(i',j')\right) \in \mathcal{E}} \exp 
\left(\dfrac{-\alpha_0 r_{i,j}}{4 \gamma_{i',j'}} \right),
\end{eqnarray}
where $\mathcal{V}_{\bfR}=\left\lbrace 1,\ldots,N_{\textrm{row}}\right \rbrace \times \left\lbrace 1,\ldots,N_{\textrm{col}}\right \rbrace$, $\mathcal{V}_{\bGam}=\left\lbrace 1,\ldots,N_{\textrm{row}}+1 \right \rbrace \times \left\lbrace 1,\ldots,N_{\textrm{col}}+1 \right \rbrace$, and the edge set $\mathcal{E}$ consists of pairs $\left((i,j),(i',j')\right)$ representing the connection between $r_{i,j}$ and $\gamma_{i',j'}$.
It can be seen from \eqref{eq:GMRF} that 
\begin{subeqnarray}
\label{eq:prior1_r}
\slabel{eq:prior1_r2}
r_{i,j}|\bGam,\alpha_0 &\sim & \mathcal{G}\left(\alpha_0, \dfrac{\alpha_{i,j}(\bGam)}{\alpha_0}\right) \\
\slabel{eq:prior1_r3}
\gamma_{i,j}|\bfR,\alpha_0 &\sim & \mathcal{IG}\left(\alpha_0,\alpha_0 \beta_{i,j}(\bfR)\right)
\end{subeqnarray}
where
\begin{eqnarray*}
\alpha_{i,j}(\bGam) & = & 4\left(\gamma_{i,j}^{-1} + \gamma_{i-1,j}^{-1} + \gamma_{i,j-1}^{-1} + \gamma_{i-1,j-1}^{-1}\right)^{-1}\\
\beta_{i,j}(\bfR) &=& \left(r_{i,j} + r_{i+1,j} + r_{i,j+1} + r_{i+1,j+1}\right)/4.
\end{eqnarray*}

Notice that we denote explicitly the dependence on the value of $\alpha_0$, which here acts a regularization parameter that controls the amount of spatial smoothness enforced by the GMRF. Following an empirical Bayesian approach, the value of $\alpha_0$ remains unspecified and will be adjusted automatically (together with the depth parameter $c$) during the inference procedure by maximum marginal likelihood estimation.

Finally, it is worth mentioning that this intensity model has similarities with the depth model \eqref{eq:MRF_depth} in the sense that spatial correlation is used to regularize the parameter estimation problem. However, the depth model \eqref{eq:MRF_depth} follows a segmentation approach in which
the target depths are assumed (and constrained) to take values in a finite set.
This leads to a piecewise constant depth representation which is usually sufficient for most applications, due to the depth resolution of the recent Lidar-based imaging systems. 

The intensity model proposed in this paper provides a spatially smooth representation of the
intensities that is possibly more realistic than a piece-wise constant representation (that would result from a TV-based intensity regularization), as it does not enforce the intensities to take a finite number of possible values.

\subsection{Prior for the background levels}
In a similar fashion to the intensities, when $r_{i,j}=0$, gamma distributions are conjugate priors for $b_{i,j}$ and the following Gamma priors 
\begin{eqnarray}
b_{i,j} \sim \mathcal{G}\left(\eta,\nu \right)
\end{eqnarray}
are assigned to $b_{i,j}$, where $\eta>0$ and $\nu>0$ are fixed hyperparameters. In order to reflect the lack of prior knowledge about the background levels, $(\eta,\nu)$ is arbitrarily set to obtain a weakly informative prior ($(\eta,\nu)=(1,10)$ in all results presented in this paper). However, $(\eta,\nu)$ can be easily adapted if additional information, e.g., observation conditions, is available. It could also be estimated as in \cite{Zhou2012}, however the choice $(\eta,\nu)=(1,10)$ made here has a limited impact on the estimation performance.

Assuming prior independence between the average background levels of the different pixels, we obtain 
\begin{eqnarray}
\label{eq:joint_prior_B}
f(\bfB|\eta,\nu) = \prod_{i,j} f(b_{i,j}|\eta,\nu).
\end{eqnarray}

\subsection{Joint posterior distribution}
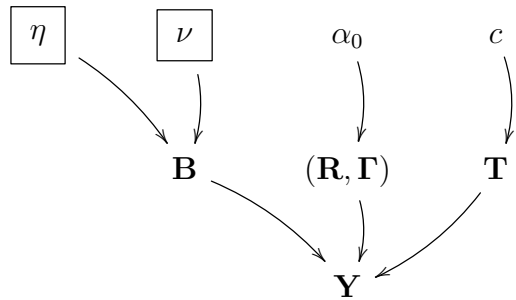
\begin{figure}[!ht]
\centerline{ \xymatrix{
 *+<0.05in>+[F-]+{\eta} \ar@/^/[rd] &  *+<0.05in>+[F-]+{\nu} \ar@/^/[d] & \alpha_0 \ar@/^/[d] 		& c \ar@/^/[d] \\
																		& \bfB \ar@/^/[rd]   								& (\bfR,\bGam) \ar@/^/[d] & \bfT \ar@/^/[ld] \\
  & & \MATpix &   }
} \caption{Graphical model for the proposed hierarchical Bayesian model (fixed quantities appear in boxes).} \label{fig:DAG}
\end{figure}

We can now specify the joint posterior distribution for $\bfT, \bfB, \bfR$ and $\bGam$ given the observed waveforms $\MATpix$ and the value of the spatial regularization parameters $c$ and $\alpha_0$ (recall that their value will be determined by maximum marginal likelihood estimation during the inference procedure). Using Bayes' theorem, and assuming prior independence between $\bfT$, $(\bGam,\bfR)$ and $\bfB$, the joint posterior distribution associated with the proposed Bayesian model is given by
\begin{eqnarray}
\label{eq:posterior}
f(&\bfT, \bfB, \bfR,\bGam|\MATpix,\alpha_3,\eta,\nu)
\propto f(\MATpix|\bfR,\bfB,\bfT)
f(\bfB|\eta,\nu)f(\bfT|c)f(\bfR,\bGam|\alpha_0).
\end{eqnarray}
For illustration, Fig. \ref{fig:DAG} depicts the directed acyclic graph (DAG) summarising the structure proposed Bayesian model (recall that $\bfR,\bGam$ have a bi-partite neighbourhood structure, which is illustrated in the graphical model of Fig. \ref{fig:neighbour_GMRF}).


\section{Bayesian Inference}\label{sec:Gibbs}
\subsection{Bayesian estimators}
\label{subsec:Bayesian_estimators}
The Bayesian model defined in Section \ref{sec:bayesian} specifies the joint posterior density for the unknown parameters $\bfT, \bfB, \bfR$ and $\bGam$ given the observed data $\MATpix$ and the parameters $c$ and $\alpha_0$. This posterior distribution models our complete knowledge about the unknowns given the observed data and the prior information available. In this section we define suitable Bayesian estimators to summarize this knowledge and perform depth imaging. More precisely, we propose to use the following two Bayesian estimators: 1) the minimum mean square error estimator (MMSE) of the intensity matrix
\begin{eqnarray}\label{reflectivityMMSE}
\hat{\bfR}_{\textrm{MMSE}} =  \textrm{E}\left[\bfR | \MATpix,\hat{c},\hat{\alpha}_0\right],
\end{eqnarray}
where the expectation is taken with respect to the marginal posterior density $f(\bfR | \MATpix,c,\alpha_0)$ (by marginalizing $\bfT, \bfB$ and $\bGam$ this density takes into account their uncertainty); 2) the maximum a posteriori (MAP) estimator of target positions
\begin{eqnarray}
\label{depthMAP}
\widehat{\bfT}_{\textrm{MAP}} = \underset{\bfT}{\textrm{argmax}}  f(\bfT | \MATpix,\hat{c},\hat{\alpha}_0),
\end{eqnarray}
which is particularly adapted to estimate discrete parameters. 
Notice that in \eqref{reflectivityMMSE} and \eqref{depthMAP}, we have set $c = \hat{c}$ and $\alpha_0 = \hat{\alpha_0}$, which denotes the maximum marginal likelihood estimator of the MRF regularisation parameters $c$ and $\alpha_0$ given the observed data $\MATpix$, i.e.,
\begin{eqnarray}\label{alphaML}
(\hat{c},\hat{\alpha}_0)=\underset{c \in \mathbb{R}^+, \alpha_0 \in \mathbb{R}^+}{\textrm{argmax}} f\left(\MATpix | c,\alpha_0\right),
\end{eqnarray}
This approach for specifying $(c,\alpha_0)$ is taken from the empirical Bayes framework in which parameters with unknown values are replaced by point estimates computed from observed data (as opposed to being fixed a priori or integrated out of the model by marginalization). As explained in \cite{Pereyra2014ssp}, this strategy has several important advantages for MRF parameters with  intractable conditional distributions such as $(c,\alpha_0)$. In particular, it allows for the automatic adjustment of the value of $(c,\alpha_0)$ for each image (thus producing significantly better estimation results than using a single fixed value of $(c,\alpha_0)$ for all data sets), and has a computational cost that is several times lower than that of competing approaches, such as including $(c,\alpha_0)$ in the model and subsequently marginalising them during the inference procedure \cite{Pereyra2013ip}.

\subsection{Bayesian algorithm}
Computing the estimators  \eqref{reflectivityMMSE} and \eqref{depthMAP} is challenging because it involves calculating expectations with respect to posterior marginal densities, which in turn require evaluating the full posterior \eqref{eq:posterior} and integrating it over a very high-dimensional space. Computing $(\hat{c},\hat{\alpha}_0)$ is also difficult because it involves solving an intractable optimisation problem, (because it is not possible to evaluate the marginal likelihood $f(\MATpix | c,\alpha_0)$ or its gradient $\nabla f(\MATpix | c,\alpha_0)$). Here we adopt the approach proposed in \cite{Pereyra2014ssp} and design a stochastic optimisation and simulation algorithm to compute \eqref{reflectivityMMSE} and \eqref{depthMAP} simultaneously. That is, we construct a stochastic gradient Markov chain Monte Carlo (SGMCMC) algorithm that simultaneously estimates $(\hat{c},\hat{\alpha}_0)$ and generates a chain of $N_{\textrm{MC}}$ samples $\{\bfR^{(n)},\bfT^{(n)},\bfB^{(n)},\bGam^{(n)}\}_{n=1}^{N_{\textrm{MC}}}$ asymptotically distributed according to the marginal density $f(\bfT, \bfB, \bfR,\bGam | \MATpix,\hat{c},\hat{\alpha}_0)$ (this algorithm is summarised in Algo. \ref{algo:algo1} below). Once the samples have been generated, the estimators \eqref{reflectivityMMSE} and \eqref{depthMAP} are approximated by Monte Carlo integration \cite[Chap. 10]{Robert2004}, i.e.,
\begin{eqnarray}
\label{eq:reflectivity_MC}
\hat{\bfR}_{MMSEj} = \dfrac{1}{N_{\textrm{MC}}-N_{\textrm{bi}}}\sum_{n=N_{\textrm{bi}}+1}^{N_{\textrm{MC}}}
	\bfR^{(n)},
\end{eqnarray}
and 
\begin{eqnarray}
\label{eq:depth_MC}
\left(\hat{t}_{i,j}\right)_{\textrm{MAPj}} = \underset{k \in {1,\ldots,T}}{\textrm{argmax}}  \sum_{n=N_{\textrm{bi}}+1}^{N_{\textrm{MC}}} \delta\left( t_{i,j}^{(n)}-k\right) 
\end{eqnarray}
where the samples from the first $N_{\textrm{bi}}$ iterations (corresponding to the transient regime or burn-in period) are discarded and where $\delta(\cdot)$ denotes the Kronecker delta function. The main steps of this algorithm are detailed in below.

\subsubsection{Sampling the target positions}
It can be seen from \eqref{eq:posterior} that \\
$f(t_{i,j}=t |\MATpix,\bfT_{\backslash i,j}, \bfB, \bfR,\bGam,c,\alpha_0,\eta,\nu)$
\begin{eqnarray}
\label{eq:post_T}
 \propto f(\Vpix{i,j}|t_{i,j}=t,r_{i,j},b_{i,j}) f(t_{i,j}=t|\bfT_{\mathcal{V}(i,j)}).
\end{eqnarray}
Consequently, sampling the target positions can be achieved by sampling sequentially each position from its conditional distribution, i.e., by drawing randomly from $\{1,\ldots,T\}$ with known probabilities. In our experiments we used a Gibbs sampler implemented using a colouring scheme such
that many positions can be updated in parallel ($9$ steps required when considering a 2-order neighborhood structure).

\subsubsection{Sampling the intensity coefficients}
Similarly, from \eqref{eq:posterior} we obtain\\
$f(\bfR |\MATpix,\bfT, \bfB,\bGam,c,\alpha_0,\eta,\nu)$
\begin{eqnarray}
\label{eq:post_R}
=\prod_{i,j}f(r_{i,j} |\Vpix{i,j},t_{i,j}, b_{i,j},\bGam,\alpha_0)
\end{eqnarray}
i.e., the elements of $\bfR$ are a posteriori independent (conditioned on the other parameters) and can thus be updated simultaneously. Moreover, \\
$f(r_{i,j} |\Vpix{i,j},t_{i,j}, b_{i,j},\bGam,\alpha_0)$
\begin{eqnarray}
\label{eq:posterior_r}
 \propto r_{i,j}^{\alpha_0-1}\exp^{-\frac{\alpha_0 r_{i,j}}{\alpha_{i,j}(\bGam)}}\prod_{\nobin=1}^{\nbbin} \lambda_{i,j,\nobin}^{\pix{i,j}{\nobin}} \exp^{-\lambda_{i,j,\nobin}},
\end{eqnarray}
with $\lambda_{i,j,\nobin}=r_{i,j}g_{0}(\nobin-\nobin_{i,j}) + b_{i,j}$. 
By noticing that $\prod_{\nobin=1}^{\nbbin} \lambda_{i,j,\nobin}^{\pix{i,j}{\nobin}}$ is a polynomial function of $r_{i,j}$, whose degree is $\sum_{t=1}^T \pix{i,j}{\nobin}$ (since $g_{0}(\nobin-\nobin_{i,j})>0, \forall t$) , it turns out that \eqref{eq:posterior_r} can be expressed as a finite mixture of gamma distributions whose weights and parameters are known (see Appendix for the derivation of \eqref{eq:posterior_r}). Sampling $r_{i,j}$ from its conditional distribution finally reduces to randomly selecting one of components of the mixture and then sampling from the corresponding gamma distribution. Note that the auxiliary variables in $\bGam$ do not appear in the likelihood \eqref{eq:likelihood} and that sampling $\bGam$ from its conditional distribution reduces to sampling from \eqref{eq:prior1_r3}.

\subsubsection{Sampling the background}
Similarly analysis applies when sampling the background levels. Precisely,\\  
$f(\bfB |\MATpix,\bfT, \bfR,\bGam,c,\alpha_0,\eta,\nu)$
\begin{eqnarray}
\label{eq:post_B}
=\prod_{i,j}f(b_{i,j} |\Vpix{i,j},t_{i,j}, r_{i,j},\eta,\nu)
\end{eqnarray}
i.e., the elements of $\bfB$ are a posteriori independent (conditioned on the other parameters) and can thus be updated simultaneously. Moreover, \\
$f(b_{i,j} |\Vpix{i,j},t_{i,j}, r_{i,j},\eta,\nu)$
\begin{eqnarray}
\label{eq:posterior_b}
 \propto b_{i,j}^{\eta-1}\exp^{-\frac{b_{i,j}}{\nu}}\prod_{\nobin=1}^{\nbbin} \lambda_{i,j,\nobin}^{\pix{i,j}{\nobin}} \exp^{-\lambda_{i,j,\nobin}},
\end{eqnarray}
which can also be expressed as a finite mixture of gamma distributions.

\subsubsection{Updating the MRF regularization parameters $c$ and $\alpha_0$}
If marginal likelihood $f(\MATpix | c,\alpha_0)$ was tractable we could update $(c,\alpha_0)$ from one MCMC iteration to the next by using a classic gradient descent step
\begin{eqnarray}
\alpha_0^{(n+1)}& =& \alpha_0^{(n)} + \xi_n \dfrac{\partial}{\partial \alpha_0} \log f(\MATpix | c^{(n)}, \alpha_0^{(n)}),\nonumber\\
c^{(n+1)}& = &c^{(n)} + \xi_n \dfrac{\partial}{\partial c} \log f(\MATpix | c^{(n)}, \alpha_0^{(n)})
\end{eqnarray}
with $\xi_n = n^{-3/4}$, such that $\alpha_0^{(n)}$ (resp. $c^{(n)}$) converges to $\hat{\alpha}_0$ (resp. $\hat{c}$) as $n \rightarrow \infty$. However, this gradient has two levels of intractability, one due to the marginalization of $(\bfT, \bfB, \bfR,\bGam)$ and another one due to the intractable normalizing constant of the gamma MRF. We address this difficulty by following the approach proposed in \cite{Pereyra2014ssp}; that is, by replacing $\nabla \log f(\MATpix | c^{(n)} \alpha_0^{(n)})$ with estimators computed with the samples generated by the MCMC algorithm at iteration $n$, and two sets of auxiliary variables. More precisely, we generate $(\bfR',\bGam') \sim \mathcal{K}_1(\bfR,\bGam|\bfR^{(n)},\bGam^{(n)},\alpha_0^{(n-1)})$ with an MCMC kernel $\mathcal{K}_1$ with target density \eqref{eq:GMRF} (in our experiments we used a Gibbs sampler implemented using a colouring scheme such that all the elements of $\bfR'$ and $\bGam'$ are generated in parallel). We also generate $\bfT' \sim \mathcal{K}_2 (\bfT|\bfT^{(n)},c^{(n-1)})$ with an MCMC kernel $\mathcal{K}_2$ with target density \eqref{eq:MRF_depth}. The updated value $\alpha_0^{(n)}$ (resp. $c^{(n)}$) is then projected onto an interval $[0,A_n]$ (resp. $[0,C_n]$, see (10:) and (11:) in Algo \ref{algo:algo1}) to guarantee the positivity constraints $c, \alpha_0 \in \mathbb{R}^+$ and the stability of the stochastic optimization algorithm (we have used $A_t = C_n =  20$).

It is worth mentioning that if it was possible to simulate the auxiliary variables $(\bfR',\bGam')$ (resp. $\bfT'$) exactly from \eqref{eq:GMRF} (resp. \eqref{eq:MRF_depth}), then the estimator of $\nabla \log f(\MATpix | c^{(n)}, \alpha_0^{(n)})$ used in Algo. \ref{algo:algo1} would be unbiased and as a result $(c^{(n)},\alpha_0^{(t)})$ would converge exactly to $(\hat{c},\hat{\alpha}_0)$. However, exact simulation from \eqref{eq:GMRF} and \eqref{eq:MRF_depth} is not computationally feasible and therefore we resort to the MCMC kernels $\mathcal{K}_1$ and $\mathcal{K}_2$ and obtain a biased estimator of $\nabla \log f(\MATpix | c^{(n)}, \alpha_0^{(n)})$ that drives $c^{(n)}, \alpha_0^{(n)}$ to a neighbourhood of $(\hat{c},\hat{\alpha}_0)$ \cite{Pereyra2014ssp}. We have found that computing this biased estimator is significantly less expensive than alternative approaches, (e.g., using an approximate Bayesian computation algorithm as in \cite{Pereyra2013ip}), and that it leads to very accurate depth and intensity results.

\section{Simulation results}
\label{sec:simulations}
\subsection{Data acquisition}
We propose comparing the performance of the proposed method to reconstruct a depth image of a life-sized polystyrene head located at a distance of $40$m from a time-of-flight scanning sensor, based on TCSPC. The transceiver system and data acquisition hardware used for this
work is broadly similar to that described in \cite{McCarthy2009,Krichel2010,Wallace2010,McCarthy2013}, which was previously developed at Heriot-Watt University. For the measurements reported in this section, the optical
path of the transceiver was configured to operate with a fibre-coupled illumination wavelength of
841 nm, and a silicon single-photon avalanche diode (SPAD) detector. The overall system had a jitter of $95$ps full width at half-maximum (FWHM).  The measurements have been performed outdoors, on the Edinburgh Campus of Heriot-Watt University, in November 2014 under dry clear skies and with atmospheric conditions remaining relatively constant
for the duration of the measurement. The key measurement parameters are summarized in Table \ref{tab:measur_param}. The acquisition time per pixel in Table \ref{tab:measur_param} is $30$ms. However, the data format of timed events allows the construction of photon timing histograms associated with shorter acquisition times, after measurement, as the system records the time of arrival of each detected photon. Here, we evaluate our algorithms for acquisition times of $30$ms, $6$ms, $3$ms, $600\mu$s, $300\mu$s and $60\mu$s per pixel.

\begin{figure}[h!]
  \centering
  \includegraphics[width=\columnwidth]{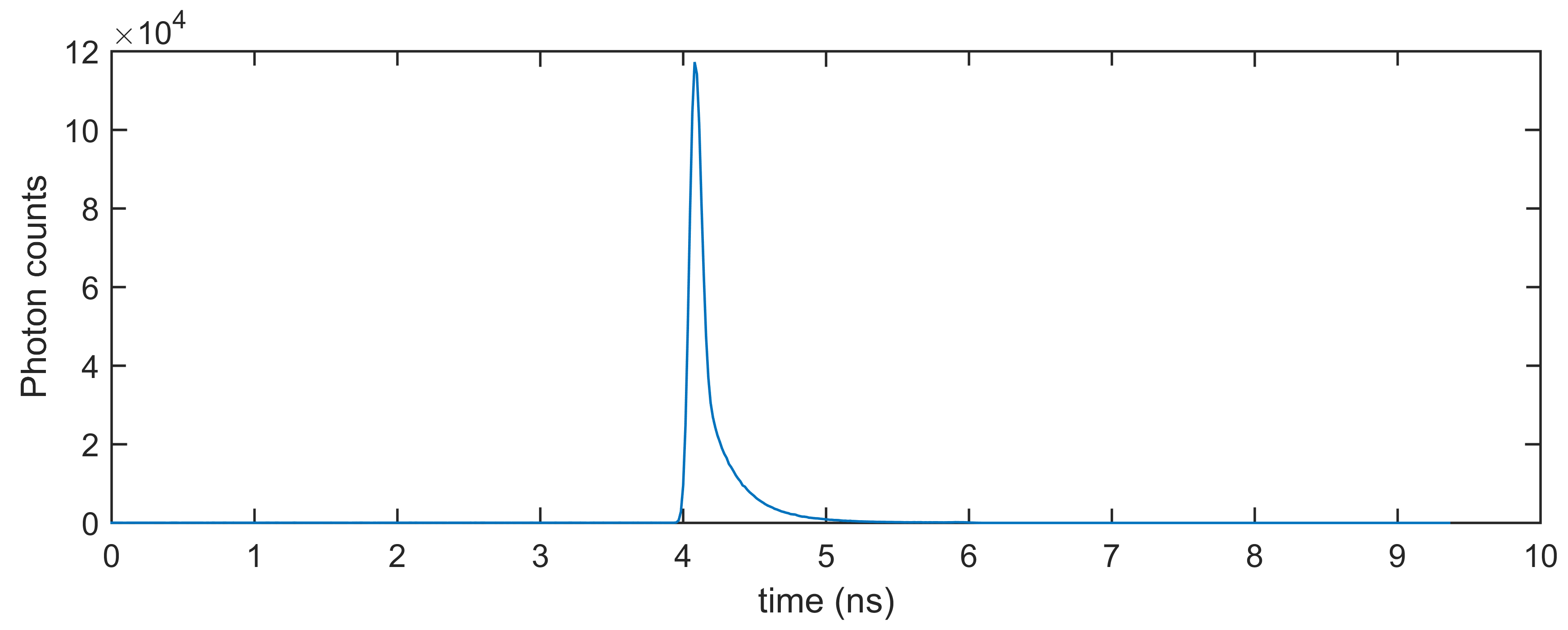}
  \caption{Intrumental response obtained using spectralon panel placed at $40$m from laser source/detector and for an acquisition time of $60$s (jitter $\approx 95$ps FWHM).}
  \label{fig:impulse_response}
\end{figure}

The instrumental impulse response $g_{0}(\cdot)$ is estimated from preliminary experiments by analysing the distribution of photons reflected onto a Spectralon panel (a commercially available Lambertian scatterer), placed at $40$m from laser source/detector. A long acquisition time ($60$s) is considered here to reduce the impact of the photon count variability and a pre-processing step is used to remove the constant background in the measured response. The resulting instrumental impulse response is depicted in Fig. \ref{fig:impulse_response}.

Table \ref{tab:data_analysis} provides details regarding the amount of detected photons when varying the acquisition time. The second row of  \ref{tab:data_analysis} shows the average number of detected photons per pixel, ranging from about $0.85$ photons for a $60\mu$s acquisition time to about $420$ for a $30$ms acquisition time. As expected, the number of detected photons increases linearly with exposure. The bottom row shows that almost $49\%$ of the pixels do not contain any detected photons for a $60\mu$s acquisition time and that this proportion decreases when increasing the acquisition time. 


\subsection{Estimation performance}
The proposed method is compared to the method classically used for depth imaging \cite{McCarthy2013} and which is divided into two steps. The first step consists of
estimating $t_{i,j}$ using cross-correlation between $g_0(\cdot)$ and the photon histogram $\Vpix{i,j}$. The object depth is estimated using 
\begin{eqnarray}
\hat{t}_{i,j,\textrm{corr}}=\underset{\tau \in \bbZ}{\textrm{argmax}} \sum_{t=1}^T \pix{i,j}{t} g_0(t-\tau).
\end{eqnarray}
Once the estimated time target distance $\hat{t}_{i,j,\textrm{corr}}$ has been computed, the target intensity
is estimated using maximum likelihood (ML) estimation (assuming that $b_{n} = 0$) as 
\begin{eqnarray}
\hat{r}_{i,j,ML}=\dfrac{\sum_{t=1}^{T} y_{i,j,t}}{\sum_{t=1}^{T} g_0\left((t-\hat{t}_{i,j,\textrm{corr}}\right)}.
\end{eqnarray} 
When the background level is relatively low compared to the maximum value of $r_{i,j} g_{0}\left((\nobin-\nobin_{i,j}\right)$, the ML intensity estimates (conditioned on the previously estimated depths) provide satisfactory results and are thus consider as the comparative method in the remainder of this paper. 
The proposed algorithm has been applied with $N_{\textrm{MC}}=1000$ iterations, including $N_{\textrm{bi}}=200$ burn-in iterations.

\begin{figure}[h!]
  \centering
  \includegraphics[width=\columnwidth]{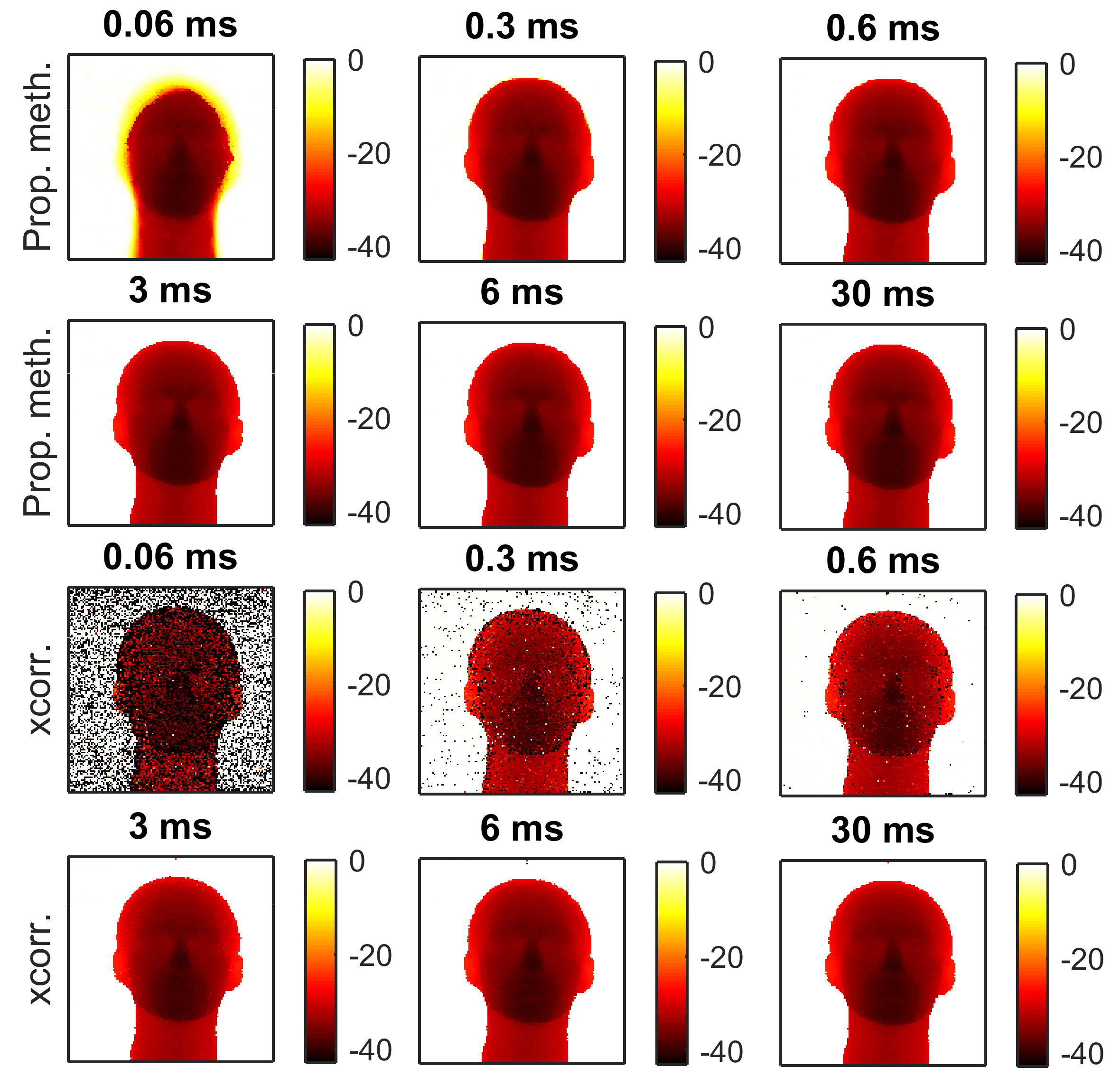}
  \caption{Depth maps for the $40$m target and for different per-pixel acquisition times, estimated by the proposed Bayesian algorithm (top rows) and the standard method (bottom rows). Distances shown are in centimeters and the reference distance is the distance of the backplane. Black pixels correspond to pixels where no photon are detected and for which the cross-correlation method cannot identify the target distance.}
  \label{fig:depth_maps}
\end{figure}

Fig. \ref{fig:depth_maps} compares the estimated depth maps obtained by the standard and the proposed methods. These results show that for large acquisition times, the two methods provide similar results. However, when the acquisition time decreases, the cross-correlation method starts to fail in identifying the target positions, especially in pixels where no photon is detected in a pixel, indicating that the proposed method seems more robust to the absence of signal in some pixels.

\begin{figure}[h!]
  \centering
  \includegraphics[width=\columnwidth]{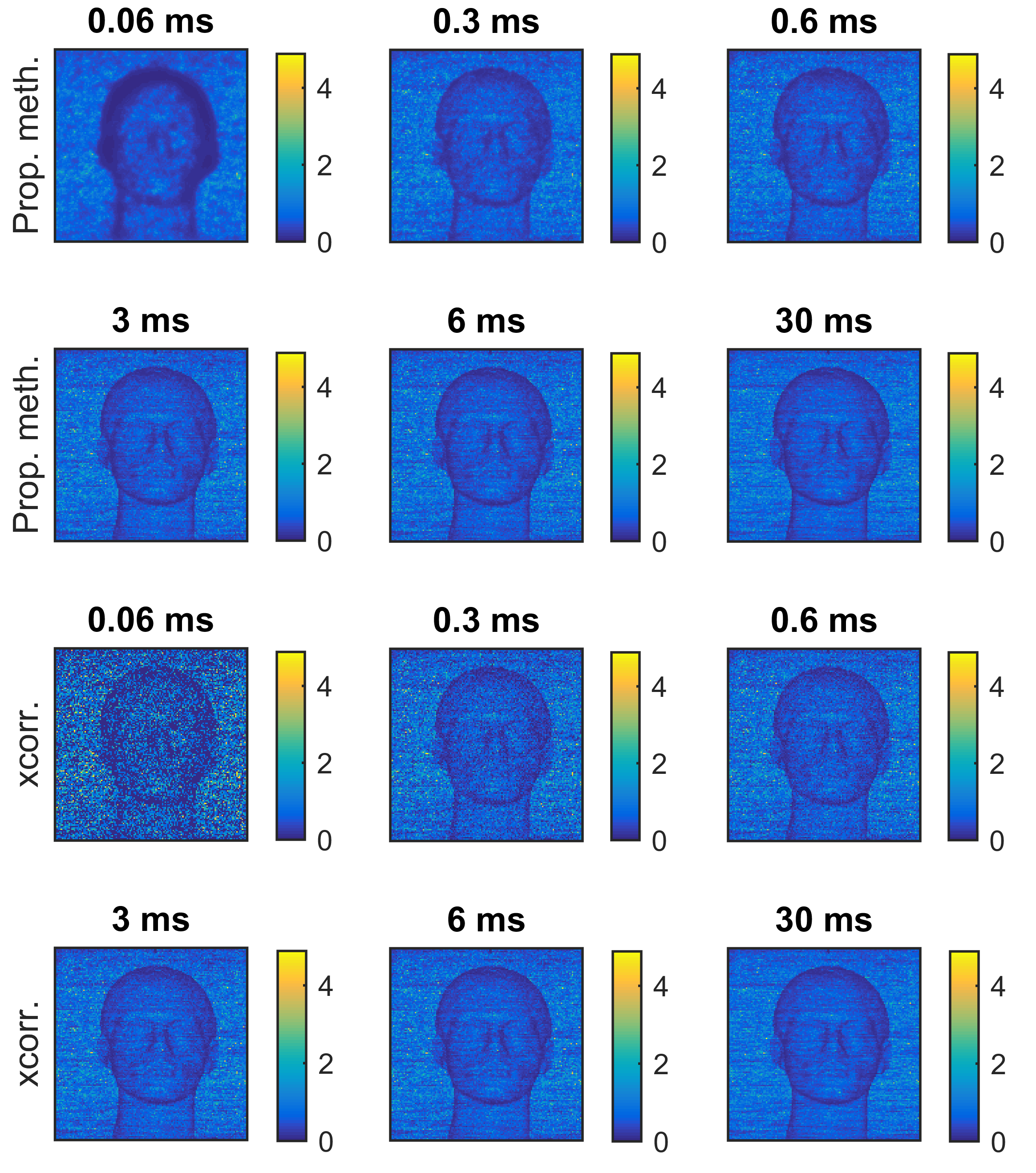}
  \caption{Intensity maps for the $40$m target and for different per-pixel acquisition times, estimated by the proposed Bayesian algorithm (top rows) and the standard method (bottom rows).}
  \label{fig:reflectivity_maps}
\end{figure}

\begin{figure}[h!]
  \centering
  \includegraphics[width=\columnwidth]{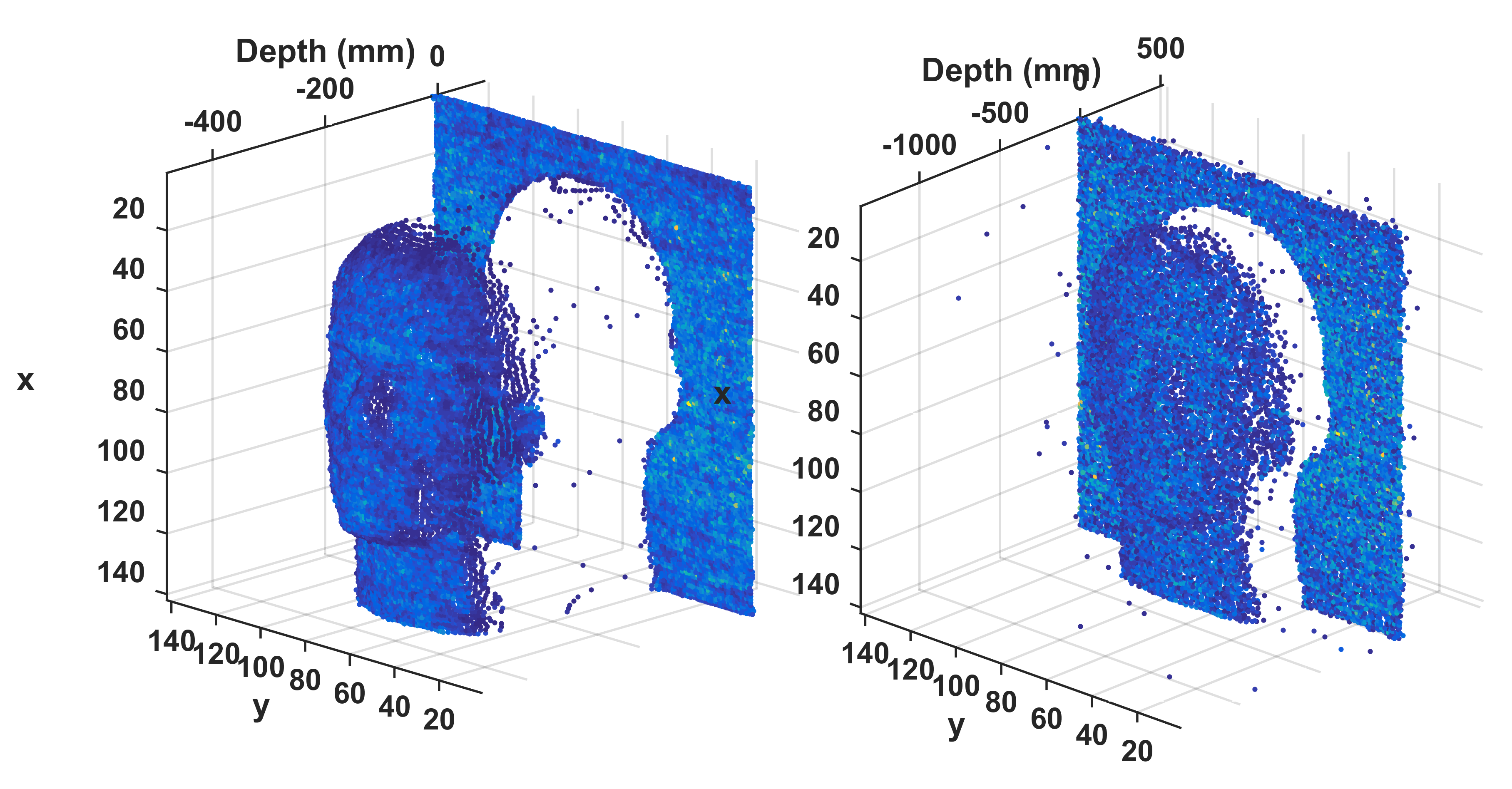}
  \caption{Depth/intensity reconstruction of the target, estimated by the Bayesian (left) and standard (right) methods for a $300\mu$s acquisition time. The colours represent the target intensity (dark blue for low intensity coefficients).}
  \label{fig:compare_300us}
\end{figure}

Fig. \ref{fig:reflectivity_maps} compares the estimated intensity maps obtained by the two methods. These results show that the two methods provide similar results for the longest acquisition times and that the proposed method is more robust to the lack/absence of detected photons. In particular, for the $60\mu$s acquisition time, few photons are detected in the pixels around the head and the proposed algorithm provides a smoother intensity image due to consideration of spatial correlations, in contrast to the standard method which process the pixels independently. Note that for each experiment, the Spectralon response $g_0(\cdot)$ is scaled to account for the acquisition time (e.g., amplitude divided by ten between the $30$ms and $3$ms experiments). Note also that for some pixels, the intensities estimated by the two methods can exceed $1$. This point will be discussed further in the conclusions. Fig. \ref{fig:compare_300us} compares the depth/intensity estimation results obtained by the two methods for an acquisition time of $300\mu$s. This figure illustrates the ability of the proposed model to handle low photon returns using the spatial correlation of the depths and intensities.

\begin{figure}[h!]
  \centering
  \includegraphics[width=\columnwidth]{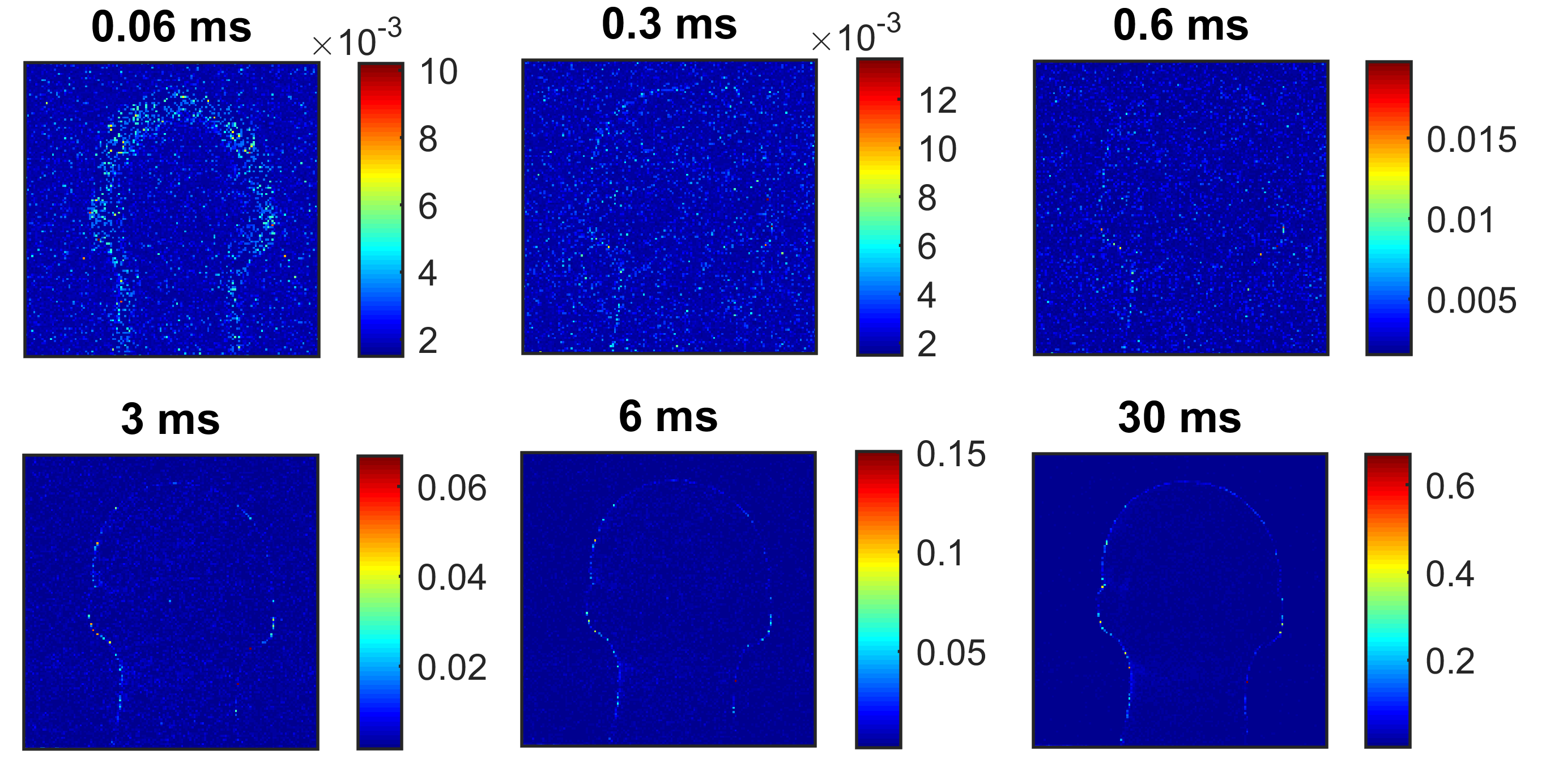}
  \caption{Background level maps for different acquisition time, estimated by the proposed Bayesian algorithm.}
  \label{fig:estimated_background}
\end{figure}

In addition to the depth and intensity maps, the proposed method also estimates the average background level in each pixel, depicted in Fig. \ref{fig:estimated_background}. This figure shows that for the longer acquisition times, higher backgrounds are estimated at the boundary between the head and the backplane, which can be explained by the presence of two peaks in the histograms of detected photons. Due to the laser beam size, some photons are reflected onto the head whereas others are reflected onto the backplane and thus arrive later onto the detector. When the number of detected photons decreases, the amplitudes of the two peaks decrease, which makes the detection of multiple peaks more difficult.

\begin{figure}[h!]
  \centering
  \includegraphics[width=\columnwidth]{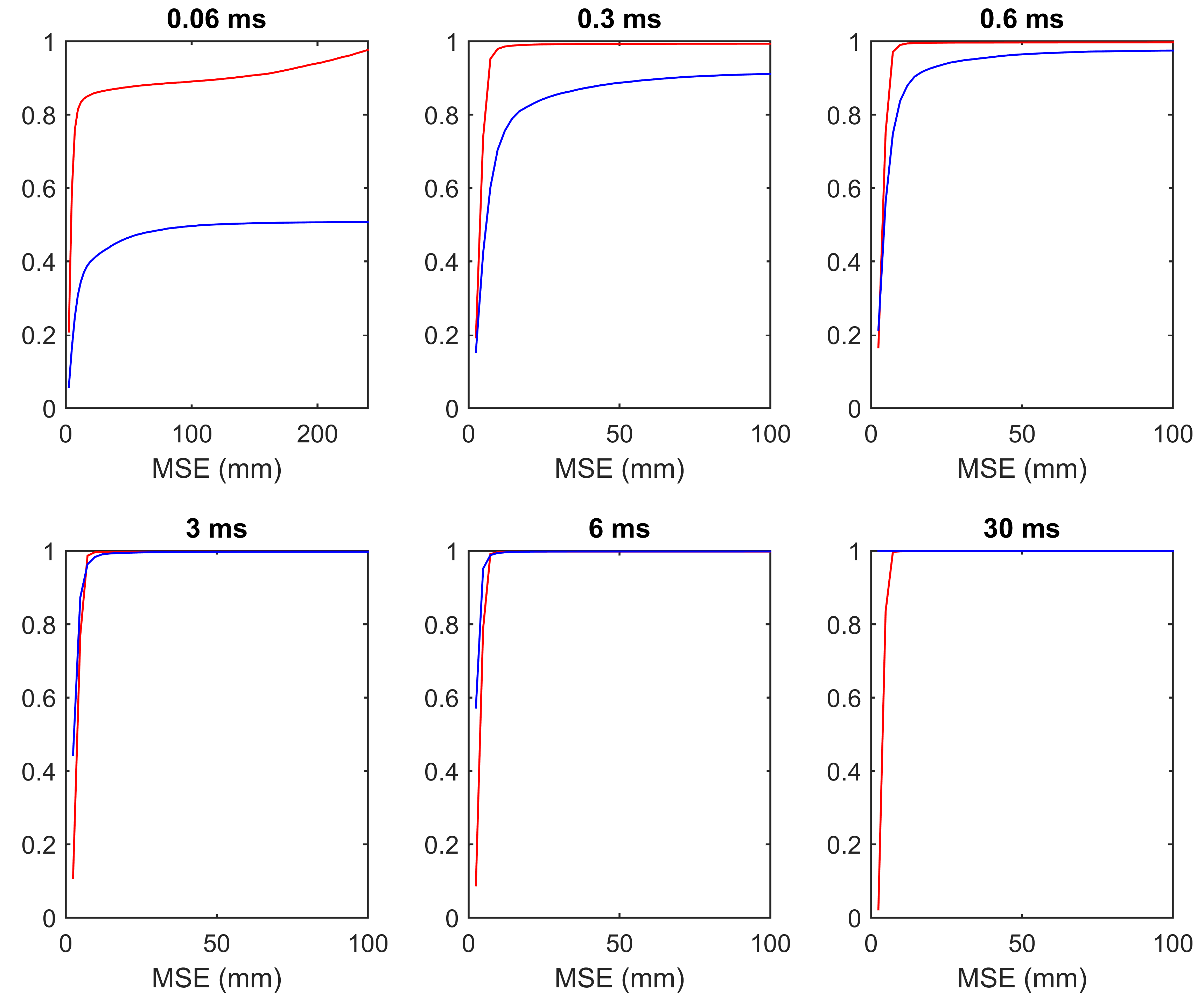}
  \caption{Distance RMSE cdfs provided by the standard (blue) and the proposed (red) methods for the target located at $40$m.}
  \label{fig:perf_depth_45m}
\end{figure}

\begin{figure}[h!]
  \centering
  \includegraphics[width=\columnwidth]{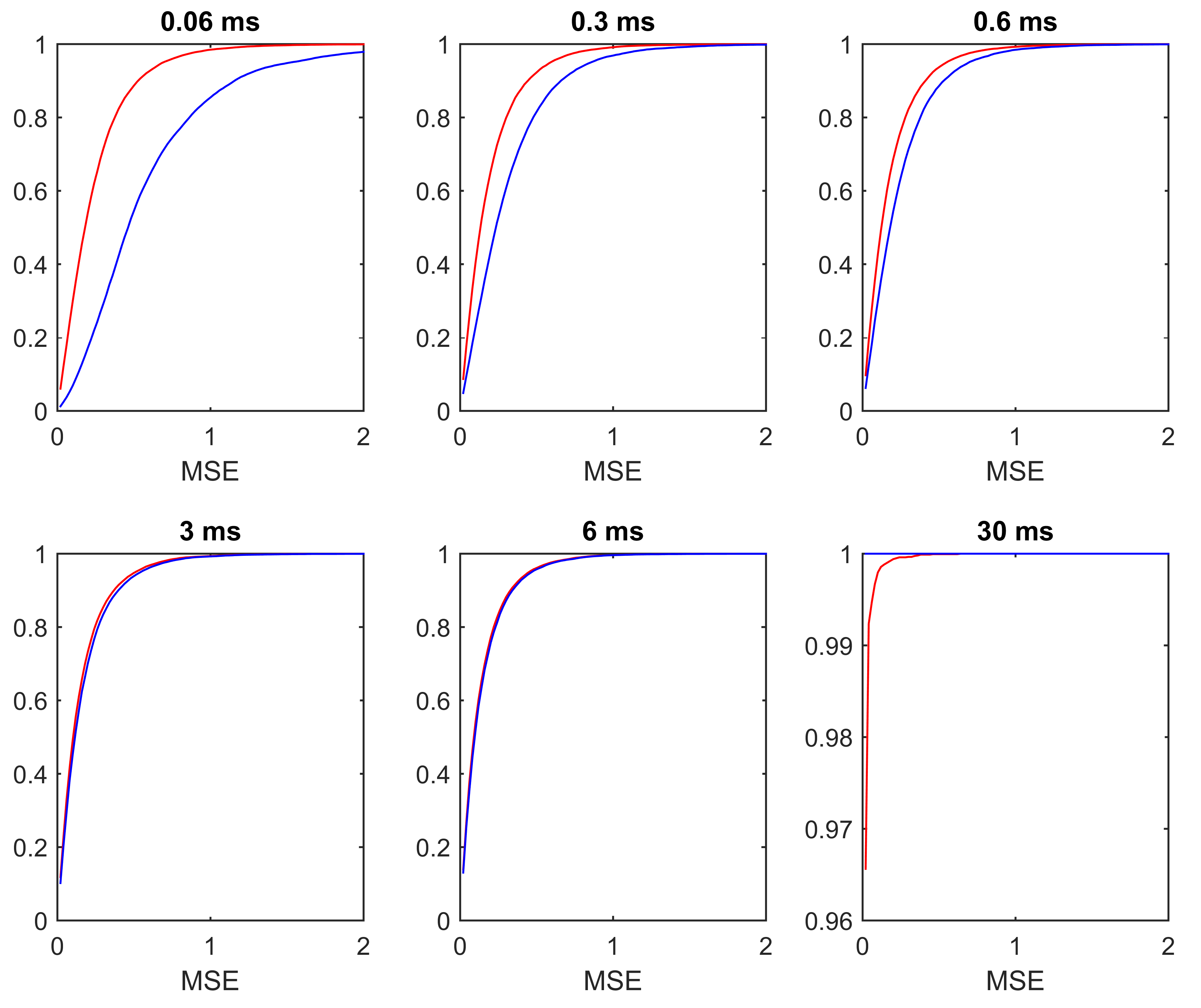}
  \caption{Intensity RMSE cdfs provided by the standard (blue) and the proposed (red) methods for the target located at $40$m.}
  \label{fig:perf_reflec_45m}
\end{figure}

The performance of the methods are quantitatively evaluated using the distance and intensity mean squared errors (MSEs) defined by 
\begin{eqnarray}
MSE(d_{i,j})=\norm{\hat{d}_{i,j}-d_{i,j}}_2^2
\end{eqnarray}
where $\norm{\cdots}_2$ denotes the $\ell_2$-norm, $\hat{d}_{i,j}$ is the estimated value of $d_{i,j}=3\times10^{8}t_{i,j}/2$ and 
\begin{eqnarray}
MSE(r_{i,j})=\norm{\hat{r}_{i,j}-r_{i,j}}_2^2
\end{eqnarray}
where $\hat{r}_{i,j}$ is the estimated value of $r_{i,j}$. 
Note that $\{d_{i,j}\}$ and $\{r_{i,j}\}$ are unknown for the data set considered. Consequently we replace these values by those estimated by the standard method for the longest acquisition time ($30$ms). Figs. \ref{fig:perf_depth_45m} and \ref{fig:perf_reflec_45m} depict the cumulative density functions (cdfs) of the distance and intensity MSEs, defined by 
\begin{eqnarray}
F_d(\tau) & = & \dfrac{1}{N_{\textrm{row}} N_{\textrm{col}}} \sum_{i,j} \Indicfun{(0,\tau)}{MSE(d_{i,j})} \\
F_r(\tau) & = & \dfrac{1}{N_{\textrm{row}} N_{\textrm{col}}} \sum_{i,j} \Indicfun{(0,\tau)}{MSE(r_{i,j})}
\end{eqnarray}
where $\Indicfun{(0,\tau)}{\cdot}$ denotes the indicator function defined on $(0,\tau)$. Figs. \ref{fig:perf_depth_45m} and \ref{fig:perf_reflec_45m} show that the proposed method is more robust than the standard method when reducing the acquisition time and provide more consistent results in terms of depth and intensity estimation. These figures also highlight the ability of the proposed method to process pixels for which no photons are detected, as the cdfs are upperbounded by the proportion of pixels that can be processed by each method. 

Finally, Table \ref{tab:computational_cost} compares the computational costs of the two methods to process the whole image ($142 \times 142$ pixels, $T=586$), for the different acquisition durations. Due to the use of the MCMC method, the proposed method is significantly more computationally demanding than the standard method. However, this cost must be balanced by the performance improvement in terms of depth and intensity estimation, when the number of detected photons is low. When the flux of detected photons is large enough, the consideration of spatial correlations has a limited impact on the estimation performance, as long as the background levels are low compared to the amplitudes of the peaks associated with actual targets. It is interesting to note the computational cost of the Bayesian increases with the acquisition time, in contrast to the standard method. This is mainly due to the MCMC steps used to updates the intensity coefficients and the background levels which require the computation of polynomial coefficients whose number depends on the number of detected photons in each pixels (see Appendix).

\section{Conclusion}
\label{sec:conclusion} 
In this paper, we proposed a new Bayesian model for Lidar-based low photon count imaging of single-layered targets. In the Bayesian framework, prior distributions were assigned to the unknown target depths and intensity to account for the intrinsic correlations between neighboring pixels. An adaptive Markov chain Monte Carlo method was then developed to estimate the model unknown parameters, including the spatial regularization parameters, thus relieving practitioners from setting these parameters by cross-validation. The model and method were validated using real Lidar data and the results showed the benefits of the proposed approach compared to the classical method used when the number of detected photons is low. 

In the paper, we assumed that the beam associated with a given pixel is incident on a single surface. This assumption is reasonable for small beam sizes, compared to the target distance and when the scene is composed of locally continuous surfaces. When the beam encounters multiple surfaces, one peak will be considered as principal surface, depending on its amplitude and on the Lidar returns in the neighboring pixels. The remaining peaks will be considered as part of the background noise. Considering returns from multiple surfaces is an interesting problem already addressed in \cite{Hernandez2007,Ramirez2012,Wallace2014} for applications where the number of detected photons are significantly higher. It would be interesting to extend this work for the low-photon imaging problem. 

Since the model considered in this paper assumed the presence of a target in each pixel, the proposed method will tend to process empty pixels (i.e., containing no photon) using the neighboring pixels, which might be inaccurate for non-locally continuous surfaces (such as wire fence). Accounting for the absence of target in some pixels is currently under investigation.

In the results presented in Section \ref{sec:simulations}, some of the estimated intensity coefficients were significantly greater than one, even for long acquisition time and even when scaling the instrumental response. Thus, the estimated intensities cannot be directly related to target reflectivity values as these estimation errors do not seem to be only due to estimation errors when extracting the instrumental impulse response. Thus constraining the intensity coefficients to be less than one might not be sufficient to provide accurate reflectivity estimates. As studied in \cite{Milonni2004,Capraro2012,henriksson2014}, atmospheric perturbations can have a significant impact on the distribution of the detected photons, specially for long-range targets. Although such scintillation effects will have a small impact on the depth estimation, accounting for them will be necessary in future work to improve the reflectivity estimation.

\section*{Appendix: On the conditional distribution of the intensity coefficients}
The conditional distribution $f(r_{i,j} |\Vpix{i,j},t_{i,j}, b_{i,j},\bGam,\alpha_0)$
can be expressed (up to a multiplicative constant) as \\
$f(r_{i,j} |\Vpix{i,j},t_{i,j}, b_{i,j},\bGam,\alpha_0)$
\begin{eqnarray}
\label{eq:post_r_2}
 \propto r_{i,j}^{\alpha_0-1}\exp^{-\frac{\alpha_0 r_{i,j}}{\alpha_{i,j}(\bGam)}}\exp^{-\sum_{\nobin=1}^{\nbbin}\lambda_{i,j,\nobin}}\prod_{\nobin=1}^{\nbbin} \lambda_{i,j,\nobin}^{\pix{i,j}{\nobin}} ,
\end{eqnarray}
where $\sum_{\nobin=1}^{\nbbin}\lambda_{i,j,\nobin}=u_{i,j}+ r_{i,j}v_{i,j}$, 
\begin{eqnarray}
u_{i,j} & = & \sum_{\nobin=1}^{\nbbin}b_{i,j}\nonumber\\
v_{i,j} & = & \sum_{\nobin=1}^{\nbbin}g_{0}(\nobin-\nobin_{i,j}),\nonumber
\end{eqnarray}
and 
\begin{eqnarray}
P(r_{i,j})=\prod_{\nobin=1}^{\nbbin} \lambda_{i,j,\nobin}^{\pix{i,j}{\nobin}}=\prod_{\nobin=1}^{\nbbin} \left(r_{i,j}g_{0}(\nobin-\nobin_{i,j}) + b_{i,j} \right)^{\pix{i,j}{\nobin}}
\end{eqnarray}
is a polynomial function of $r_{i,j}$. Since $g_{0}(\nobin-\nobin_{i,j})>0, \forall \nobin$, $-b_{i,j}/g_{0}(\nobin-\nobin_{i,j})$ is a root of $P(r_{i,j})$ if $\pix{i,j}{\nobin}>0$. Moreover, this root is of multiplicity $\pix{i,j}{\nobin}$ and the polynomial order is thus $O_{i,j}=\sum_{t=1}^T \pix{i,j}{\nobin}$. Let 
\begin{eqnarray}
P(r_{i,j})=\sum_{k=0}^{O_{i,j}} \epsilon_k r_{i,j}^k,
\end{eqnarray}
be the polynomial expansion of $P(r_{i,j})$, whose coefficients $\{\epsilon_k\}$ can be obtained from the polynomial roots. From Eq. \eqref{eq:post_r_2}, we obtain\\
$f(r_{i,j} |\Vpix{i,j},t_{i,j}, b_{i,j},\bGam,\alpha_0)$
\begin{eqnarray}
\propto \sum_{k=0}^{O_{i,j}} \epsilon_k r_{i,j}^{\alpha_0+k-1}\exp^{-r_{i,j} \left(\frac{\alpha_0}{\alpha_{i,j}(\bGam)}+v_{i,j}\right)}.
\end{eqnarray}
which can be expressed as the following mixture of $O_{i,j}+1$ gamma distributions
\begin{eqnarray}
f(r_{i,j} |\Vpix{i,j},t_{i,j}, b_{i,j},\bGam,\alpha_0) = \sum_{k=0}^{O_{i,j}} w_{k} \mathcal{G}\left(r_{i,j};\alpha_0+k,\left(\frac{\alpha_0}{\alpha_{i,j}(\bGam)}+v_{i,j}\right)^{-1} \right),
\end{eqnarray}
with
\begin{eqnarray}
w_{k} \propto  \epsilon_k \dfrac{\Gamma\left(\alpha_0+k\right)}{\left(\frac{\alpha_0}{\alpha_{i,j}(\bGam)}+v_{i,j}\right)^{\alpha_0+k}},\quad \forall k,
\end{eqnarray}
where $\Gamma(\cdot)$ denotes the Gamma function
and $\sum_{k=0}^{O_{i,j}}w_{k}=1$.
\bibliographystyle{IEEEtran}
\bibliography{biblio}

\clearpage

\begin{algogo}{Proposed MCMC algorithm}
     \label{algo:algo1}
     \begin{algorithmic}[1]
        \STATE \underline{Fixed input parameters:} Lidar impulse response $g_0(\cdot)$, number of burn-in iterations $N_{\textrm{bi}}$, total number of iterations $N_{\textrm{MC}}$
				\STATE \underline{Initialization ($n=0$)}
        \begin{itemize}
        \item Set $\bfR^{(0)},\bfT^{(0)},\bfB^{(0)},\bGam^{(0)},c^{(0)},\alpha_0^{(0)}$
        \end{itemize}
        \STATE \underline{Iterations ($1 \leq n \leq N_{\textrm{MC}}$)}
        \STATE Sample $\bfT^{(n)}$ from \eqref{eq:post_T}
        \STATE Sample $\bfR^{(n)}$ from \eqref{eq:post_R}
				\STATE Sample $\bfB^{(n)}$ from \eqref{eq:post_B}
				\STATE Sample $\bGam^{(n)}$ from \eqref{eq:prior1_r3}
				\IF{$n<N_{\textrm{bi}}$} 
				\STATE Sample $(\bfR',\bGam') \sim \mathcal{K}_1(\bfR,\bGam|\bfR^{(n)},\bGam^{(n)},\alpha_0^{(n-1)})$
				\STATE Sample $\bfT' \sim \mathcal{K}_2(\bfT|\bfT^{(n)},c^{(n-1)})$
				\STATE Set $\alpha_0^{(n)} = \mathcal{P}_{[0,A_n]}\left(\alpha_0^{(n-1)} + \xi_n \left[\Lambda(\bfR^{(n)},\bGam^{(n)})-\Lambda(\bfR',\bGam') \right]\right)$\\
				where $\Lambda(\bfR,\bGam)=\sum_{(i,j) 
\in \mathcal{V}_{\bfR}} \log(r_{i,j}) -\sum_{(i',j') \in \mathcal{V}_{\bGam}}\log(\gamma_{i',j'}) - \sum_{\left((i,j),(i',j')\right) \in \mathcal{E}}\dfrac{r_{i,j}}{4 \gamma_{i',j'}}$
				\STATE Set $c^{(n)} = \mathcal{P}_{[0,C_n]}\left(c^{(n-1)} + \xi_n \left[\phi(\bfT')-\phi(\bfT^{(n)}) \right]\right)$
				\ELSE
				\STATE Set $c^{(n)} = c^{(n-1)}$
				\STATE Set $\alpha_0^{(t)} = \alpha_0^{(n-1)}$
				\ENDIF
        \STATE Set $n = n+1$.
        \STATE Output $\{\bfR^{(n)},\bfT^{(n)} \}_{n=1}^{N_{\textrm{MC}}}$.
        \end{algorithmic}
\end{algogo}

\begin{figure}[h!]
  \centering
  \includegraphics[width=\columnwidth]{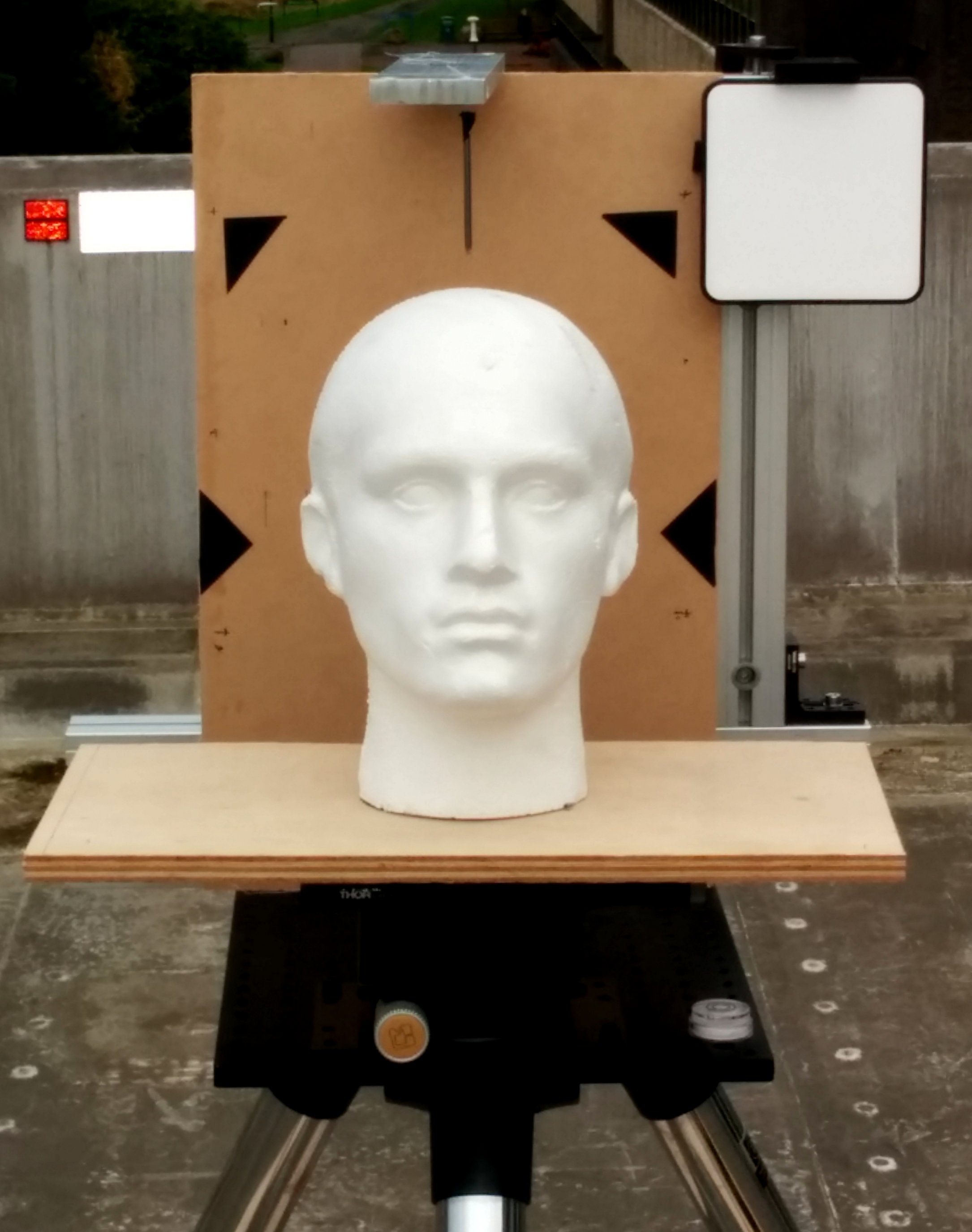}
  \caption{Photograph showing the polystyrene head used for the experiments
described here and calibration targets, including the Spectralon panel (top right corner).}
  \label{fig:photo_setup}
\end{figure}

\begin{table}[h!]
\renewcommand{\arraystretch}{1.2}
\begin{footnotesize}
\begin{center}
\begin{tabular}{|c|c|}
\hline
Target Stand-off Distance & $\approx 40$m\\
\hline
\multirow{7}{*}{Target Scene} & Polystyrene head \\
& ($\approx 170 \times 285 \times
250$mm\\
& in $W \times H \times D$ when\\
&  viewed from the front)\\
& mounted on a breadboard. \\
& Backplane: MDF board.\\
& (see Fig \ref{fig:photo_setup})\\
\hline
\multirow{6}{*}{Laser system} & Supercontinuum \\
& laser source and\\
&  tunable filter \\
& (NKT Photonics)\\
& fibre-coupled to the \\
& custom-designed transceiver unit\\
\hline
Illum. Wavelength & $841$nm\\
\hline
Laser Repetition Rate & $19.5$MHz\\
\hline
Illum. Power at target & $\approx 240\mu$W average optical power\\
\hline
Illum. Beam Diameter at Target & $\approx 1$mm\\
\hline
\multirow{6}{*}{Acquisition Mode} & $142 \times 142$ pixels scan \\
& centred on the head,\\
& covering an area of \\
& $285 \times 285$mm at the scene\\
& Per-pixel acquisition time: $30$ ms\\
& Total scan time: $\approx 10$ minutes\\
\hline
Histogram bin width & $16$ps\\
\hline
Histogram length &  586 bins (after gating)\\
\hline
Temporal Response of System & $\approx 95$ps FWHM\\
\hline
\end{tabular}
\end{center}
\end{footnotesize}
\caption{Measurement key parameters.\label{tab:measur_param}}
\end{table}

\begin{table}[h!]
\renewcommand{\arraystretch}{1.2}
\begin{footnotesize}
\begin{center}
\begin{tabular}{|c|c|c|c|c|c|c|}
\cline{2-7}
\multicolumn{1}{c|}{} & $60\mu$s & $300\mu$s & $600\mu$s & $3$ms & $6$ms & $30$ms \\
\hline
Av. Photon counts & $0.8$ & $4.2$ & $8.4$ & $42.0$ & $83.7$ & $418.6$ \\
\hline
Empty pixels ($\%$) &$48.7$ & $7.5$& $1.7$ & $\geq 0.1$ & $0$ & $0$ \\
\hline
\end{tabular}
\end{center}
\end{footnotesize}
\caption{Average number of detected photons per pixel and proportion of empty pixels as a function of the acquisition time.\label{tab:data_analysis}}
\end{table}

\begin{table}[h!]
\renewcommand{\arraystretch}{1.2}
\begin{footnotesize}
\begin{center}
\begin{tabular}{|c|c|c|c|c|c|c|}
\cline{2-7}
\multicolumn{1}{c|}{} & $60\mu$s & $300\mu$s & $600\mu$s & $3$ms & $6$ms & $30$ms \\
\hline
X-corr+MLE & $1$ & $1$ & $1$ & $1$ & $1$ & $1$ \\
\hline
Prop. method &$113$ & $123$& $131$ & $152$ & $197$ & $347$ \\
\hline
\end{tabular}
\end{center}
\end{footnotesize}
\caption{Processing time (in minutes).\label{tab:computational_cost}}
\end{table}

\end{document}

%% file: notations_yoann.tex
\input com_notations.tex

\newcommand{\Vpix}[1]{\mathbf{y}_{#1}}

\newcommand{\MATpix}{\mathbf{Y}}
\newcommand{\pix}[2]{y_{#1,#2}}

\newcommand{\nbbin}{T}
\newcommand{\nobin}{t}

\newcommand{\transp}{^T}

\newcommand{\norm}[1]{\left\|#1\right\|}

\newcommand{\Indicfun}[2]{\textbf{1}_{#1}\left(#2\right)}

\newenvironment{algogo}[1]{
\smallskip
\noindent \hrule\vspace{0.2\baselineskip} \hrule
\begin{small}
\refstepcounter{algo} \center{\bf \textsc{Algorithm \thealgo}}
\\{\center{\bf #1}}
\smallskip
\flushleft
 } {
\end{small}
\smallskip
\hrule\vspace{0.2\baselineskip} \hrule
\smallskip
}

\newcounter{algo}
\renewcommand{\thealgo}{\arabic{algo}}

%% file: com_notations.tex
\def\bfB{{\mathbf{B}}}

\def\bfR{{\mathbf{R}}}

\def\bfT{{\mathbf{T}}}

\def\bbZ{{\mathbb{Z}}}
